\documentclass[preprint,review,12pt]{elsarticle}

\usepackage{hyperref}

\usepackage{amssymb}

\newtheorem{theorem}{Theorem}[section]

\journal{Chaos, Solitons \& Fractals}

\usepackage{graphicx}
\usepackage{mathptmx}      
\usepackage{amssymb}
\usepackage[utf8]{inputenc}
\usepackage{bm}
\newcommand{\be}{\begin{equation}}
\newcommand{\ee}{\end{equation}}

\begin{document}

\begin{frontmatter}

\title{Bifurcations and transition to chaos in generalized fractional maps of the orders $0<\alpha<1$}





%

\author[mymainaddress,mysecondaryaddress]{Mark Edelman\corref{mycorrespondingauthor}}
\cortext[mycorrespondingauthor]{Corresponding author}
\ead{edelman@cims.nyu.edu}
\author[thirdaddress] {Avigayil B. Helman}  
\ead{avigayilhelman34@gmail.com}
\author[fourthaddress]{Rasa Smidtaite} 
\ead{rasa.smidtaite@ktu.lt}

\address[mymainaddress]{Stern College for Women, Yeshiva University, 245 Lexington Ave., New York, NY 10016, USA}
\address[mysecondaryaddress]{Courant Institute of Mathematical Sciences at NYU, 251 Mercer Street, New York, NY 10012, USA}
\address[thirdaddress]{School of Engineering and Applied Sciences, Columbia University, 500 W. 120th Street, New York, NY 10027, USA}
\address[fourthaddress]{Department of Applied Mathematics, Kaunas University of Technology, Studentu 50-429, 51368, Kaunas, Lithuania}

\begin{abstract}
Generalized fractional maps of the orders $0<\alpha<1$ are Volterra difference equations of convolution type with kernels, which differences are absolutely summable, but the series of kernels are diverging. Commonly used in applications fractional (with the power-law kernels) and fractional difference (with the falling factorial kernels) maps of the orders $0<\alpha<1$ belong to this class of maps. We derived the algebraic equations which allow the calculation of bifurcation points of generalized fractional maps. We calculated the bifurcation points and drew bifurcation diagrams for the fractional and fractional difference logistic maps. Although the transition to chaos on individual trajectories in fractional maps may be characterized by the cascade of bifurcations type behavior and zero Lyapunov exponents, 
the results of our numerical simulations allow us to make a conjecture that the cascade of bifurcations scenarios of transition to chaos in generalized fractional maps and regular maps are similar, and 
the value of the generalized fractional Feigenbaum constant $\delta_f$ is the same as the value of the regular Feigenbaum constant $\delta=4.669...$.
\end{abstract}


\end{frontmatter}


\section{Introduction}
\label{sec:1}

Introduced in \cite{TarZas2008} for orders $\alpha>1$ and extended in 
\cite{MEChaos2013} to $\alpha>0$ fractional and proposed in \cite{Fall,Chaos2014} fractional difference maps are Volterra difference equations of convolution type with the power-law (in fractional maps) and falling factorial (in fractional difference maps) kernels. These maps were recently used in cryptography \cite{Crypto2016} and biological \cite{ME11}, economic \cite{TarEc2}, communication \cite{Rag}, and control (multiple papers) applications. It is the researchers' task to find which maps to use in a particular application or which one is a better fit for a natural phenomenon. Fractional and fractional difference maps share many features like cascade of bifurcations type trajectories and power-law convergence to the fixed points (see, e.g. reviews      \cite{HBV2,HBV4}). Their properties to a high degree are defined by the common properties of their kernels which belong to a class of discrete functions whose series are diverging; however, the series of their differences belong to the space of absolutely converging series $l_1$. For this reason, the generalized fractional maps, which include fractional and fractional difference maps, were introduced in \cite{ME14,Helman}.

The universal cascade of bifurcations scenario of transition to chaos is   well-known (the most popular reference is Robert May's ``Nature'' article \cite{May}). The foundational papers on universality in non-linear dynamics are collected in the reprinted selection of articles \cite{Cvi}. 
Mathematically, the universality in regular nonlinear dynamics is expressed through the Feigenbaum function and constants (see, e.g., \cite{Fei1,Fei2}), and is based on the analysis of periodic points and their bifurcations. Fractional systems have no bifurcation points \cite{PerD1,PerD2,PerC1,PerC2,PerC3,PerC4,PerC5}, and the corresponding analysis cannot be directly extended to the maps with the asymptotically power-law memory.

Although the fractional maps do not have periodic points, they have 
asymptotically periodic points. Asymptotically periodic points and stability of the fixed points in generalized fractional maps were investigated in the recent papers \cite{ME14,Helman,SF,FDperBif}. 
In this paper, we use the results of \cite{ME14} to draw the bifurcation diagrams for the fractional difference logistic map.
We also derive the equations which define bifurcation points in generalized fractional maps. We use them to calculate the bifurcation points in fractional and fractional difference logistic maps. The problem that we are trying to address is whether the universal behavior persists in generalized fractional maps. Our numerical results suggest that the answer is ``yes'' and that the fractional Feigenbaum constant $\delta$ has the same value as the regular Feigenbaum constant.

We performed numerical simulations of the fractional and fractional difference logistic maps for various values of the order $\alpha \in (0,1)$. In this paper, we present only selected numerical results for the order $\alpha=0.5$ (the value which is exactly in the middle of the interval between two integer values) to demonstrate our findings, which are valid for all considered values of $\alpha$.

\section{Preliminaries}
\label{sec:2}

The generalized universal $\alpha$-family of maps of the orders $0<\alpha<1$ is defined as (see \cite{ME14,Helman}):
\begin{eqnarray}
x_{n}= x_0 
-\sum^{n-1}_{k=0} G^0(x_k) U_\alpha(n-k),
\label{FrUUMapN}
\end{eqnarray} 
where $G^0(x)=h^\alpha G_K(x)/\Gamma(\alpha)$, $x_0$ is the initial condition, $h$ is the time step of the map, $\alpha$ is the order of the map, $G_K(x)$ is a nonlinear function depending on the parameter $K$, $U_\alpha(n)=0$ for $n \le 0$, and $U_\alpha(n) \in \mathbb{D}^0(\mathbb{N}_1)$. The space $\mathbb{D}^i(\mathbb{N}_1)$ is defined as (see \cite{Helman})
{\setlength\arraycolsep{0.5pt}
\begin{eqnarray}
&&\mathbb{D}^i(\mathbb{N}_1) \ \ = \ \ \{f: |\sum^{\infty}_{k=1}\Delta^if(k)|>N, 
\nonumber \\
&& \forall N, \ \ N \in \mathbb{N}, 
\sum^{\infty}_{k=1}|\Delta^{i+1}f(k)|=C, \ \ C \in \mathbb{R}_+\},
\label{DefForm}
\end{eqnarray}
}
where the forward difference operator $\Delta$ is defined as
\begin{equation}
\Delta f(n)= f(n+1)-f(n).
\label{Delta}
\end{equation}
For any real number $a$, $\mathbb{N}_a=\{a, a+1, a+2, a+3, ...\}$, and $\mathbb{N}=\mathbb{N}_0$. 
To generalize a one-dimensional map 
\begin{equation}
x_{n+1}= F_K(x_n)
\label{1Dmap}
\end{equation}
function $G_K(x)$ is defined as 
\begin{equation}
G_K(x_n)=x_{n}-F_K(x_n)
\label{1DmapFrGen}
\end{equation}
to converge to the map Eq.~(\ref{1Dmap})
when $\alpha=1$, $h=1$, and $U_1(n)=1$.

In the case of Caputo fractional maps \cite{TarZas2008,MEChaos2013}, $U_{\alpha}(n)$ is a power function
\begin{equation}
U_{\alpha}(n)=n^{\alpha-1}.
\label{1DmapFrGenN}
\end{equation}

In the case Caputo fractional difference maps, which are defined as solutions of the Caputo h-difference equation \cite{Fall,Chaos2014,DifSum}
\begin{equation}
(_0\Delta^{\alpha}_{h,*} x)(t) = -G_K(x(t+(\alpha-1)h)),
\label{LemmaDif_n_h}
\end{equation}
where $t\in (h\mathbb{N})_{m}$, with the initial conditions 
 \begin{equation}
(_0\Delta^{k}_h x)(0) = c_k, \ \ \ k=0, 1, ..., m-1, \ \ \ 
m=\lceil \alpha \rceil,
\label{LemmaDifICn_h}
\end{equation}
the kernel $U_\alpha(n)$ is the falling factorial function: 
{\setlength\arraycolsep{0.5pt}
\begin{eqnarray}
&&U_{\alpha}(n)=(n+\alpha-2)^{(\alpha-1)} 
, \   \ U_{\alpha}(1)=(\alpha-1)^{(\alpha-1)}=\Gamma(\alpha).
\label{UnFrDif}
\end{eqnarray} 
}
The falling factorial $t^{(\alpha)}$ is defined as
\begin{equation}
t^{(\alpha)} =\frac{\Gamma(t+1)}{\Gamma(t+1-\alpha)}, \ \ t\ne -1, -2, -3.
...
\label{FrFacN}
\end{equation}
The falling factorial is asymptotically a power function (See Eq.~(32) and Fig. 4 from \cite{Chaos2014}):
\begin{equation}
\lim_{t \rightarrow
  \infty}\frac{\Gamma(t+1)}{\Gamma(t+1-\alpha)t^{\alpha}}=1,  
\ \ \ \alpha \in  \mathbb{R}.
\label{GammaLimitN}
\end{equation}
The $h$-falling factorial $t^{(\alpha)}_h$ is defined as \cite{h_fac} 
\begin{eqnarray}
&&t^{(\alpha)}_h =h^{\alpha}\frac{\Gamma(\frac{t}{h}+1)}{\Gamma(\frac{t}{h}+1-\alpha)}= h^{\alpha}\Bigl(\frac{t}{h}\Bigr)^{(\alpha)}, \  \
\frac{t}{h} \ne -1, -2, -3,
....
\label{hFrFacN}
\end{eqnarray}
 
The asymptotically period-$l$ points in generalized fractional maps are defined by the following equations \cite{ME14}
{\setlength\arraycolsep{0.5pt}   
\begin{eqnarray} 
&&x_{l,m+1}-x_{l,m}=S_{1,l} G^0(x_{l,m})+\sum^{m-1}_{j=1}S_{j+1,l} G^0(x_{l,m-j})
\nonumber \\
&&+\sum^{l-1}_{j=m}S_{j+1,l} G^0(x_{l,m-j+l}), \  \ 0<m<l,
\label{LimDifferences}
\\
&&\sum^{l}_{j=1} G^0(x_{l,j})=0,
\label{LimDifferencesN}
\end{eqnarray}
}
where
{\setlength\arraycolsep{0.5pt}   
\begin{eqnarray} 
&&S_{j+1,l}=\sum^{\infty}_{k=0}\Bigl[
U_{\alpha} (lk+j) - U_{\alpha} (lk+j+1)\Bigr], \  \ 0 \le j<l.
\label{Ser}
\end{eqnarray}
}
In this paper we will use the following easily verifiable identity:
\begin{equation}
\sum^{l}_{j=1}S_{j,l}=0.
\label{Ssum}
\end{equation}


\section{Equations defining $T=2^{n-1}$ --  $T=2^{n}$ bifurcation points}
\label{sec:3}

Let us consider a cycle with the period $l=2^n$.
We expect that near the $T=2^{n-1}$ -- $T=2^{n}$ bifurcation point the difference $x_{2^n,i+2^{n-1}}-x_{2^n,i}$ will be small. To obtain the value of this difference, we perform the summation of  Eq.~(\ref{LimDifferences}) from $m=i$ to $m=i+2^{n-1}-1$ and use the results of the following transformations:
{\setlength\arraycolsep{0.5pt}   
\begin{eqnarray} 
&&\sum^{i+2^{n-1}-1}_{m=i}(x_{2^n,m+1}-x_{2^n,m})= x_{2^n,i+2^{n-1}}-x_{2^n,i}; 
\label{Sum1}
\\
&&\sum^{i+2^{n-1}-1}_{m=i}\sum^{m-1}_{j=1}S_{j+1,2^n} G^0(x_{2^n,m-j})=
\sum^{i+2^{n-1}-1}_{m=i}\sum^{m-1}_{j=1}S_{m-j+1,2^n} G^0(x_{2^n,j})
\nonumber \\
&&=\sum^{i-1}_{j=1}G^0(x_{2^n,j})\sum^{i+2^{n-1}-1}_{m=i}S_{m-j+1,2^n} 
+\sum^{i+2^{n-1}-2}_{j=i}G^0(x_{2^n,j})\sum^{i+2^{n-1}-1}_{m=j+1}S_{m-j+1,2^n};
\label{Sum2}
\\
&&\sum^{i+2^{n-1}-1}_{m=i}\sum^{2^n-1}_{j=m}S_{j+1,2^n} G^0(x_{2^n,m-j+2^n})=\sum^{i+2^{n-1}-1}_{m=i}\sum^{2^n}_{j=m+1} S_{m-j+2^n+1,2^n} G^0(x_{2^n,j})
\nonumber \\
&&=\sum^{i+2^{n-1}}_{j=i+1}G^0(x_{2^n,j})\sum^{j-1}_{m=i}S_{m-j+2^n+1,2^n}
\nonumber \\
&&+\sum^{2^n}_{j=i+2^{n-1}+1}G^0(x_{2^n,j})\sum^{i+2^{n-1}-1}_{m=i}S_{m-j+2^n+1,2^n};
\label{Sum3} \\
&&0<i\le 2^{n-1}.
\nonumber 
\end{eqnarray}
}
The resulting set of equations can be written as
{\setlength\arraycolsep{0.5pt}   
\begin{eqnarray} 
&&x_{2^n,i+2^{n-1}}-x_{2^n,i}= 
\sum^{i+2^{n-1}-1}_{j=i}G^0(x_{2^n,j})S_{1,2^n}+
\sum^{i-1}_{j=1}G^0(x_{2^n,j})\sum^{i+2^{n-1}-1}_{m=i}S_{m-j+1,2^n} 
\nonumber \\
&&+\sum^{i+2^{n-1}-2}_{j=i}G^0(x_{2^n,j})\sum^{i+2^{n-1}-1}_{m=j+1}S_{m-j+1,2^n}+
\sum^{i+2^{n-1}}_{j=i+1}G^0(x_{2^n,j})\sum^{j-1}_{m=i}S_{m-j+2^n+1,2^n}
\nonumber \\
&&+\sum^{2^n}_{j=i+2^{n-1}+1}G^0(x_{2^n,j})\sum^{i+2^{n-1}-1}_{m=i}S_{m-j+2^n+1,2^n};
\label{SumAll} \\
&&0<i\le 2^{n-1}.
\nonumber 
\end{eqnarray}
}
It is easy to see that this formula can be written as 
{\setlength\arraycolsep{0.5pt}   
\begin{eqnarray} 
&&x_{2^n,i+2^{n-1}}-x_{2^n,i}= 
\sum^{2^{n-1}}_{j=1}[G^0(x_{2^n,j})-G^0(x_{2^n,2^{n-1}+j})]\sum^{i+2^{n-1}-1}_{m=i}S_{m-j+1,2^n}; 
\label{SumAllShort} \\
&&0<i\le 2^{n-1},
\nonumber 
\end{eqnarray}
}
where coefficients $S_{k,2^n}$, defined by Eq.~(\ref{Ser}) for $0<k \le 2^n$, are continued periodically for $\forall k, \ \ k \in \mathbb{Z}$:
{\setlength\arraycolsep{0.5pt}   
\begin{eqnarray} 
&&S_{k,2^n}=S_{k+2^n,2^n}.
\label{SPer} \\
\nonumber 
\end{eqnarray}
} 
Indeed, let us consider the terms on RHS of Eq.~(\ref{SumAll}) containing the same factor $G^0(x_{2^n,p})$. Only the second sum contains the terms with $0<p<i$. They all can be written as
{\setlength\arraycolsep{0.5pt}   
\begin{eqnarray} 
&&G^0(x_{2^n,p})\sum^{i+2^{n-1}-1}_{m=i}S_{m-p+1,2^n}. 
\label{p_lt_i} 
\end{eqnarray}
}
Only the first and the third sums contain the terms with $p=i$:
{\setlength\arraycolsep{0.5pt}   
\begin{eqnarray} 
&&G^0(x_{2^n,p})\Bigl[S_1+\sum^{i+2^{n-1}-1}_{m=i+1}S_{m-p+1,2^n}\Bigr]= G^0(x_{2^n,p})\sum^{i+2^{n-1}-1}_{m=i}S_{m-p+1,2^n}. 
\label{p_eq_i} 
\end{eqnarray}
}
The first, the third, and the fourth sums contain the terms with $i<p \le i+2^{n-1}-2$:
 {\setlength\arraycolsep{0.5pt}   
\begin{eqnarray} 
&&G^0(x_{2^n,p})\Bigl[S_1+\sum^{i+2^{n-1}-1}_{m=p+1}S_{m-p+1,2^n}
+\sum^{p-1}_{m=i}S_{m-p+1,2^n}\Bigr]
\nonumber \\
&&= G^0(x_{2^n,p})\sum^{i+2^{n-1}-1}_{m=i}S_{m-p+1,2^n}. 
\label{ThirdInt} 
\end{eqnarray}
}
Only the first and the fourth sums contain the terms with            $p=i+2^{n-1}-1$:
{\setlength\arraycolsep{0.5pt}   
\begin{eqnarray} 
&&G^0(x_{2^n,p})\Bigl[S_1+\sum^{i+2^{n-1}-2}_{m=i}S_{m-p+1,2^n}\Bigr]= G^0(x_{2^n,p})\sum^{i+2^{n-1}-1}_{m=i}S_{m-p+1,2^n}. 
\label{FourthInt} 
\end{eqnarray}
}
It is obvious that for $p=i+2^{n-1}$ the fourth sum and for $i+2^{n-1}<p \le 2^{n}$ the fifth sum produce the same formula Eq.~(\ref{p_lt_i}). Now we may write
{\setlength\arraycolsep{0.5pt}   
\begin{eqnarray} 
&&x_{2^n,i+2^{n-1}}-x_{2^n,i}= 
\sum^{2^{n-1}}_{j=1}G^0(x_{2^n,j})\sum^{i+2^{n-1}-1}_{m=i}S_{m-j+1,2^n}
\nonumber  \\
&&+\sum^{2^{n}}_{j=2^{n-1}+1}G^0(x_{2^n,j})\sum^{i+2^{n-1}-1}_{m=i}S_{m-j+1,2^n};  \ \
0<i\le 2^{n-1}.
\label{Total1}
\end{eqnarray}
}
After substitutions $p=j-2^{n-1}$ and $q=m-2^{n-1}$, the second sum in the last expression can be written as
{\setlength\arraycolsep{0.5pt}   
\begin{eqnarray} 
&&\sum^{2^{n}}_{j=2^{n-1}+1}G^0(x_{2^n,j})\sum^{i+2^{n-1}-1}_{m=i}S_{m-j+1,2^n}=\sum^{2^{n-1}}_{p=1}G^0(x_{2^n,2^{n-1}+p})\sum^{i-1}_{q=i-2^{n-1}}S_{q-p+1,2^n}
\nonumber \\
&&=\sum^{2^{n-1}}_{p=1}G^0(x_{2^n,2^{n-1}+p})\Bigl[\sum^{i+2^{n-1}-1}_{q=i-2^{n-1}}S_{q-p+1,2^n}-\sum^{i+2^{n-1}-1}_{q=i}S_{q-p+1,2^n}\Bigr]
\nonumber \\
&&=-\sum^{2^{n-1}}_{j=1}G^0(x_{2^n,2^{n-1}+j})\sum^{i+2^{n-1}-1}_{m=i}S_{m-j+1,2^n};  \ \
0<i\le 2^{n-1}.
\label{Total2}
\end{eqnarray}
}
Validity of Eq.~(\ref{SumAllShort}) follows from Eqs.~(\ref{Total1})~and~(\ref{Total2}).

Near the $T=2^{n-1}$ --  $T=2^{n}$ bifurcation point, $x_{2^n,j+2^{n-1}} \approx x_{2^n,j} \approx  x_{2^{n-1}bif,j}$ and
the differences $x_{2^n,j+2^{n-1}}-x_{2^n,j}$ are small. The set $\{ x_{2^{n-1}bif,1},x_{2^{n-1}bif,2}, . . ., x_{2^{n-1}bif,2^{n-1}}\}$ is the $2^{n-1}$-cycle at the bifurcation point. In the linear approximation, we may write 
{\setlength\arraycolsep{0.5pt}   
\begin{eqnarray} 
&&G^0(x_{2^n,j+2^{n-1}})-G^0(x_{2^n,j})=\frac{dG^0(x)}{dx}\Bigl|_{x_{2^{n-1}bif,j}}( x_{2^n,j+2^{n-1}}-x_{2^n,j}). 
\label{DifBif1} 
\end{eqnarray}
}    
Near the bifurcation point, Eq.~(\ref{SumAllShort}) can be written as
{\setlength\arraycolsep{0.5pt}   
\begin{eqnarray} 
&&\sum^{2^{n-1}}_{j=1}\Bigl[\frac{dG^0(x)}{dx}\Bigl|_{x_{2^{n-1}bif,j}}
 \sum^{i+2^{n-1}-1}_{m=i}S_{m-j+1,2^n}+\delta_{i,j}\Bigr]( x_{2^n,j+2^{n-1}}-x_{2^n,j})=0; 
\label{BifEq} \\
&&0<i\le 2^{n-1},
\nonumber 
\end{eqnarray}
}
where $ \delta_{i,j}$ is the Kronecker symbol. System of $2^{n-1}$ linear equations Eq.~(\ref{BifEq}) has a nonzero solution only if the determinant of $2^{n-1}$ by $2^{n-1}$ matrix $A$ of its coefficients
{\setlength\arraycolsep{0.5pt}   
\begin{eqnarray} 
&&A_{i,j}=\frac{dG^0(x)}{dx}\Bigl|_{x_{2^{n-1}bif,j}}
 \sum^{i+2^{n-1}-1}_{m=i}S_{m-j+1,2^n}+\delta_{i,j}
\label{Det} 
\end{eqnarray}
}
is equal to zero:
{\setlength\arraycolsep{0.5pt}   
\begin{eqnarray} 
&&\det(A)=0
\label{Det1}.
\end{eqnarray}
}

The results of this section may be summarized as a theorem: 
\begin{theorem}\label{Th1} 
The $T=2^{n-1}$ --  $T=2^{n}$ bifurcation points, $2^{n-1}$ values of $x_{2^{n-1}bif,i}$ with $0<i \le 2^{n-1}$ and the value of the nonlinear parameter $K_{2^{n-1}bif}$, of a fractional generalization of a nonlinear one-dimensional map $x_{n+1}=F_K(x_n)$ written as the Volterra difference equations of convolution type
\begin{eqnarray}
x_{n}= x_0 
-\sum^{n-1}_{k=0} G^0(x_k) U_\alpha(n-k),
\label{FrUUMapN1}
\end{eqnarray} 
where $G^0(x)=h^\alpha G_K(x)/\Gamma(\alpha)$, $x_0$ is the initial condition, $h$ is the time step of the map, $\alpha$ is the order of the map, $G_K(x)=x-F_K(x)$, $U_\alpha(n)=0$ for $n \le 0$, $U_\alpha(n) \in \mathbb{D}^0(\mathbb{N}_1)$, and
{\setlength\arraycolsep{0.5pt}
\begin{eqnarray}
&&\mathbb{D}^i(\mathbb{N}_1) \ \ = \ \ \{f: |\sum^{\infty}_{k=1}\Delta^if(k)|>N, 
\nonumber \\
&& \forall N, \ \ N \in \mathbb{N}, 
\sum^{\infty}_{k=1}|\Delta^{i+1}f(k)|=C, \ \ C \in \mathbb{R}_+\},
\label{DefForm1}
\end{eqnarray}
}
are defined by the system of $2^{n-1}+1$ equations 
{\setlength\arraycolsep{0.5pt}   
\begin{eqnarray} 
&&x_{2^{n-1}bif,m+1}-x_{2^{n-1}bif,m}=S_{1, 2^{n-1}}                 G^0(x_{2^{n-1}bif,m})+\sum^{m-1}_{j=1}S_{j+1,2^{n-1}} G^0(x_{2^{n-1}bif,m-j})
\nonumber \\
&&+\sum^{2^{n-1}-1}_{j=m}S_{j+1,2^{n-1}} G^0(x_{2^{n-1}bif,m-j+2^{n-1}}), \  \ 0<m<2^{n-1},
\label{LimDifferencesNN}
\\
&&\sum^{2^{n-1}}_{j=1} G^0(x_{2^{n-1}bif,j})=0,
\label{LimDifferencesNN2}
\\
&&\det(A)=0,
\label{DetNN}
\end{eqnarray}
}
where
{\setlength\arraycolsep{0.5pt}   
\begin{eqnarray} 
&&S_{j+1,l}=\sum^{\infty}_{k=0}\Bigl[
U_{\alpha} (lk+j) - U_{\alpha} (lk+j+1)\Bigr], \  \ 0 \le j<l, 
\nonumber \\
&&S_{i,l}=S_{i+l,l}, \ \ i \in \mathbb{Z},
\label{SerNN1}
\end{eqnarray}}
and the elements of the $2^{n-1}$-dimensional matrix $A$ are 
{\setlength\arraycolsep{0.5pt}   
\begin{eqnarray} 
&&A_{i,j}=\frac{dG^0(x)}{dx}\Bigl|_{x_{2^{n-1}bif,j}}
 \sum^{i+2^{n-1}-1}_{m=i}S_{m-j+1,2^n}+\delta_{i,j} \  \ 0 < i,j \le 2^{n-1}.
\label{DetNN2} 
\end{eqnarray}
}

\end{theorem}

\section{Bifurcation diagrams and transition to chaos in the fractional and fractional difference logistic maps with $0<\alpha<1$}
\label{sec:4}

\begin{figure}[!t]
\includegraphics[width=1.0 \textwidth]{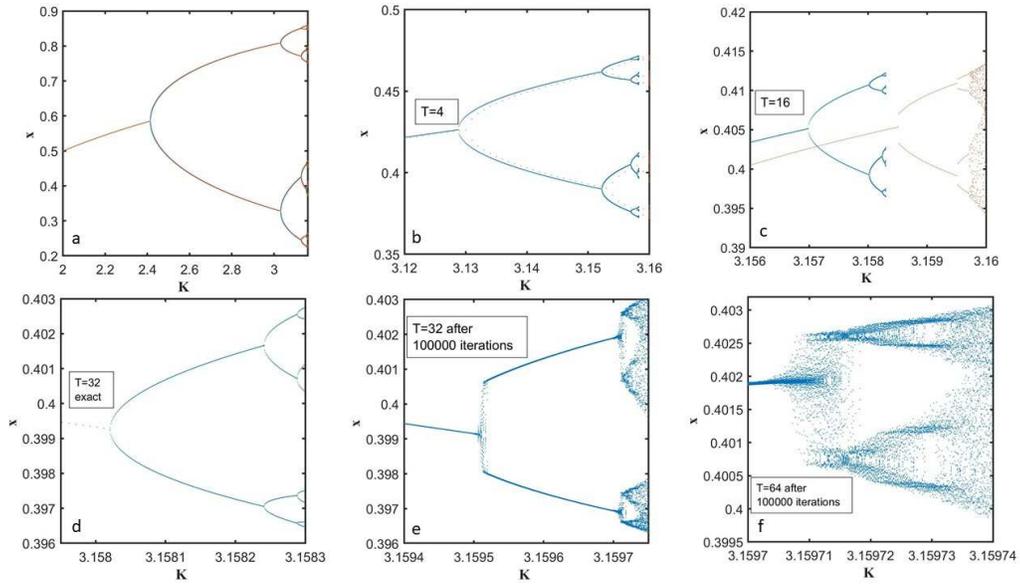}
\vspace{-0.25cm}
\caption{A part of the bifurcation diagram for the Caputo fractional difference logistic map of the order $\alpha=0.5$ from $K=2$ (fixed point) to approximately $K=3.16$ (T=128 periodic point). In figures a-c the steady line represents the solution of  Eqs.~(\ref{LimDifferences})~and~(\ref{LimDifferencesN}) (the exact solution) and the dots represent numerical calculations on a single trajectory with $x_0=0.3$ after $10^5$ iterations. 
Figure d represents the exact solution from $T=32$ on the left to $T=128$ on the right. Figures e and f represent the calculations on a single trajectory.
}
\label{fig1}
\end{figure}

In the generalized fractional logistic map 
\begin{eqnarray}
G_K(x)=x-Kx(1-x).
\label{LM}
\end{eqnarray} 
We solved Eqs.~(\ref{LimDifferences})~and~(\ref{LimDifferencesN}) numerically using the Mathematica's Newton's method (the exact solution) to draw the bifurcation diagrams for the fractional difference and fractional logistic maps with various values of $\alpha \in (0,1)$ (we assume $h=1$). Fig.~\ref{fig1} shows the results of our calculations for the fractional difference logistic map with $\alpha=0.5$ (the value which is far from the integer values one and zero).
From Fig.~\ref{fig1}b one may see that already for the period four ($T=4$) points there is a noticeable difference between the exact solutions and the results obtained after $10^5$ iterations on a single trajectory. For $T>16$ (Figs.~\ref{fig1}~b,~e,~f) the results obtained after $10^5$ iterations on a single trajectory seem to be noticeably inaccurate. Multiple papers investigating various fractional difference maps contain bifurcation diagrams obtained by iterations on a single trajectory and the number of iterations in all these papers is much less than $10^5$.
\begin{figure}[!t]
\includegraphics[width=.9 \textwidth]{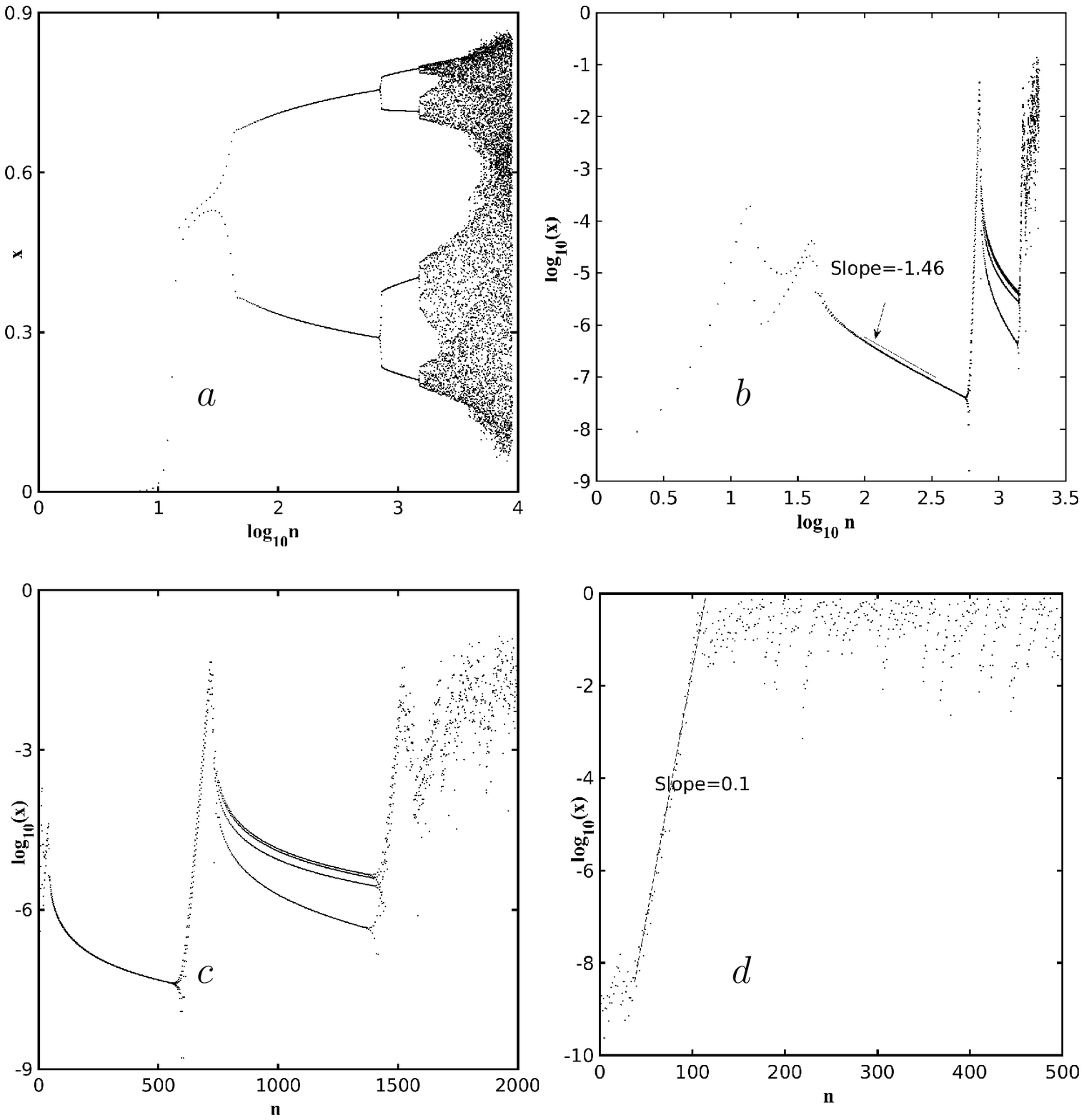}
\vspace{-0.25cm}
\caption{Transition to chaos in the Caputo fractional difference logistic map of the order $\alpha=0.2$ with $K=3.3$. a. The trajectory $x=x_n(n)$ with the initial condition $x_0=10^{-6}$; b. log-log plot of the deviation of two trajectories $x=x_{n,2}-x_{n,1}$ with the initial conditions  $x_{0,1}=10^{-6}$ and $x_{0,2}=10^{-6}+10^{-9}$; c. $\log_{10}(x)$ vs. $n$ plot for the case in b; d. logarithm of deviation vs. $n$ plot for the trajectories with $x_{0,1}=0.3$ and $x_{0,2}=0.3+10^{-9}$.
}
\label{fig2}
\end{figure} 

Bifurcation diagrams for the fractional and fractional difference logistic maps with various values of $0<\alpha<1$ are qualitatively the same. Bifurcation diagrams obtained with the increasing number of iterations on a single trajectory (and the exact bifurcation diagrams when the number of iterations goes to infinity), in many cases, demonstrate the increasing shift to the left (see Figs.~\ref{fig1}~b~and~\ref{fig1}~c). This phenomenon was first noticed and explained on the examples of the fractional and fractional difference standard maps in \cite{Chaos2014} (see Fig.~8 there). The explanation of the left shift is based on another phenomenon typical for fractional and fractional difference maps, cascade of bifurcations type trajectories (CBTT), first noticed in \cite{ME2} on the example of the order $\alpha=1.65$ standard map (Fig.~5 in that article). CBTT appear in higher periodicity asymptotically stable trajectories which first converge to a fixed point and then evolve following the cascade of bifurcations type scenario converging to an asymptotically periodic cycle (see, e.g., Figs.~6.6a~and~6.7a from \cite{ME11}). Even chaotic trajectories may first converge to a fixed point or a periodic cycle (see Fig.~7 from \cite{Chaos2014} or Fig.~3 from \cite{HBV4}). In fractional difference maps, convergence of trajectories to fixed points follows the power law
\cite{Anh} and the corresponding Lyapunov exponents are equal to zero. In the case of chaotic trajectories, the rate of convergence depends on the initial conditions. Two initially close trajectories in the asymptotically chaotic fractional difference logistic map ($\alpha=0.2$ and $K=3.3$) Fig.~\ref{fig2}a, depending on the initial conditions, may diverge exponentially Fig.~\ref{fig2}d and the corresponding Lyapunov exponent is positive. But the graph of the deviation of two initially close to zero trajectories Fig.~\ref{fig2}b~and~Fig.~\ref{fig2}c shows the transition to chaos through a series of power-law convergencies followed by jumps. 

Fractional and fractional difference maps may also demonstrate inverse CBTT, when an asymptotically periodic trajectory starts as a chaotic or higher periodicity trajectory. Examples of the inverse CBTT in the fractional and fractional difference logistic maps are Figs.~7~and~8 from \cite{MEChaos2013} and Figs.~7.6b~and~6.7c from \cite{ME11}. The inverse CBTT will cause the right shift in bifurcation diagrams. Whether the asymptotic bifurcation diagram of a fractional/fractional difference map is shifted to the left or to the right with respect to the bifurcation diagram obtained after a finite number of iterations on single trajectories depends on the type of the map and the initial conditions used.

\begin{table}[ht!]
\centering
    \begin{tabular}{| c  | c   |  c  |  c    |      c  | c      | c       |}
    \hline 
    $n $ & $K_1(n)$ & $K_{.5}(n)$ & $\frac{\Delta K_{1}(n-2)}{\Delta K_{1}(n-1)}$  &  $\frac{\Delta K_{.5}(n-2)}{\Delta K_{.5}(n-1)}$ & $\delta_1-\delta_F$ & $\delta_{.5}-\delta_F$  \\ \hline
    1          & 3         & 2.41421356 &          &          &         & \\ \hline
    2          & 3.4494897 & 3.03150807 &          &          &         & \\ \hline
    3          & 3.5440904 & 3.12880294 & 4.751446 & 6.3446 &$8.2\times10^{-2}$ & 1.6754 \\ \hline
    4          & 3.5644073 & 3.15225457 & 4.656251 & 4.1487 &$-1.3\times10^{-2}$ &
-.5205  \\ \hline
    5          & 3.5687594 & 3.15698186 & 4.668242 & 4.9609 &$-9.6\times10^{-4}$ &     .2917  \\ \hline
    6          & 3.5696916 & 3.15802086 & 4.668739 & 4.5499 &$-4.6\times10^{-4}$ & -.1193  \\ \hline
    7          & 3.5698913 & 3.15824198 & 4.669132 & 4.7199 &$-6.9\times10^{-5}$ & .0507  \\ \hline
    8          & 3.5699340 & 3.15828833 & 4.669183 & 4.6496 &$-1.9\times10^{-5}$ & -.0196 \\ \hline  
    9          & 3.5699432 & 3.15829845 & 4.669195 & 4.6766 &$-7\times10^{-6}$ & .0074  \\ \hline
    10         & 3.5699451 & 3.15830062 &          & 4.6665 &         & -.0027 \\ \hline
     11         &           & 3.158301084 &          & 4.6702 &         &   .0010 \\ \hline
     12         &           & 3.158301183 &          & 4.6688 &         & -.0004 \\ \hline
   \end{tabular}
    \caption{Approaching the Feigenbaum constant $\delta$ in the regular and fractional difference logistic maps. $K_1(n)$ are the values of the map parameter for the period $2^{n-1}$ -- period $2^{n}$ bifurcation points in the regular logistic map ($\alpha=1$); $K_{.5}(n)$ are the same points in the case $\alpha=.5$; $\delta_1=\Delta K_{1}(n-2)/\Delta K_{1}(n-1)= [K_{1}(n-1)-K_1(n-2)]/[K_{1}(n)-K_1(n-1)]$; $\delta_{.5}$ is the same value in the case $\alpha=.5$; $\delta_F=\delta=4.6692016$.}
    \label{table:T3_Fr}
\end{table}
Table~\ref{table:T3_Fr} shows the values of the map parameter for the period $2^{n-1}$ -- period $2^{n}$ bifurcation points $K_1(n)$ (for the regular logistic map) and $K_{.5}(n)$ (for the order $\alpha=0.5$ fractional difference logistic map). The results for $n=1,2,...,10$ and $\alpha=0.5$ were obtained using Eqs.~(\ref{LimDifferences})~and~(\ref{LimDifferencesN}). 
The table also shows the corresponding values of the ratios $\delta_1=\Delta K_1(n-2)/\Delta K_1(n-1)= [K_1(n-1)-K_1(n-2)]/[K_1(n)-K_1(n-1)]$ and $\delta_{.5}=\Delta K_{.5}(n-2)/\Delta K_{.5}(n-1)= [K_{.5}(n-1)-K_{.5}(n-2)]/[K_{.5}(n)-K_{.5}(n-1)]$. All results for $\alpha=0.5$ in Table~\ref{table:T3_Fr} were later calculated using Theorem \ref{Th1} (Eqs.~(\ref{LimDifferencesNN})-(\ref{DetNN})), and for the first ten values of $n$ the results are identical. Numerical calculations were performed using Matlab and the obtained results had higher accuracy than the corresponding solutions of Eqs.~(\ref{LimDifferences})~and~(\ref{LimDifferencesN}) obtained using Mathematica.
The values for the regular logistic map are available from many sources  (see, e.g., \cite{FC,FC1}. The number of significant digits in the tables is dictated by the requirement that the table should fit the standard journal's page. Higher accuracy results for the case $\alpha=1$ are available at multiple sites on the Internet. 
Corresponding results for the order $\alpha=0.5$ fractional logistic map are presented in Table~\ref{table:T3_FrN}.
\begin{table}[ht!]
\centering
    \begin{tabular}{| c  | c   |  c      | c       |}
    \hline 
    $n (T=2^n)$ & $K_{.5}(n)$  &  $\frac{\Delta K_{.5}(n-2)}{\Delta K_{.5}(n-1)}$ & $\delta_{.5}-\delta_F$  \\ \hline
    1          & 3.93016668194129 &          &          \\ \hline
    2          & 4.88614847910753 &          &          \\ \hline
    3          & 5.05712582113625 & 5.591278 & 0.922076\\ \hline
    4          & 5.09634066404542 & 4.360016 & -.309186\\ \hline
    5          & 5.10446168317469 & 4.828808 & 0.159606\\ \hline
    6          & 5.10622604348767 & 4.602812 & -.066389\\ \hline
    7          & 5.10660170543380 & 4.696670 & 0.027469\\ \hline
    8          & 5.10668234310063 & 4.658641 & -.010561\\ \hline
    9          & 5.10669959857451 & 4.673164 & 0.003963\\ \hline
    10         & 5.10670329532456 & 4.667742 & -.001460\\ \hline
   \end{tabular}
    \caption{The order $\alpha=0.5$ fractional logistic map:  the values of the map parameter for the bifurcation points $K_{.5}$ and the ratios $\delta_{.5}=\Delta K_{.5}(n-2)/\Delta K_{.5}(n-1)= [K_{.5}(n-1)-K_{.5}(n-2)]/[K_{.5}(n)-K_{.5}(n-1)]$
converging to the Feigenbaum number $\delta_F=\delta=4.6692016$.}
    \label{table:T3_FrN}
\end{table}

As in the case of the regular logistic map, the values of $\delta_{.5}$ oscillate around the Feigenbaum number but converge significantly slower. The slow convergence is expected because, in many cases, the convergence in fractional maps follows the power law while the convergence in regular maps is exponential. From the authors' point of view, the data present sufficient evidence to make a conjecture that the Feigenbaum number $\delta$ exists in fractional and fractional difference maps and has the same value as in regular maps.

\section{Conclusion}
\label{sec:5}
In this paper, we derived the analytic expressions for the coefficients of the equations that define bifurcation points in generalized fractional maps of the orders $0<\alpha<1$ (Eqs.~(\ref{LimDifferencesNN})--(\ref{DetNN2})). They allow calculations of asymptotic bifurcation points in various cases which include fractional and fractional difference maps and maps. We showed that derived in \cite{ME14} equations, which define asymptotically periodic points, Eqs.~(\ref{LimDifferences})~and~(\ref{LimDifferencesN}), allow drawing of bifurcation diagrams which are much more accurate than bifurcation diagrams obtained by iterations on single trajectories. 

Based on the data from Tables~\ref{table:T3_Fr}~and~\ref{table:T3_FrN}, we made a conjecture that the Feigenbaum number $\delta$ exists in fractional and fractional difference maps and has the same value as in regular maps. This should be a result of the symmetry and self-similarity
in the system of equations Eqs.~(\ref{LimDifferencesNN})--(\ref{DetNN2}). 
We hope that the following mathematical analysis of the equations will lead to the explanation of this symmetry.

\section*{Data availability}

Data will be made available on request.

\section*{Acknowledgements}
The first author acknowledges support from Yeshiva University's 2021--2022 Faculty Research Fund, expresses his gratitude to the administration of Courant Institute of Mathematical Sciences at NYU
for the opportunity to perform some of the computations at Courant,
and expresses his gratitude to Virginia Donnelly for technical help.

The last author acknowledges support from the European Social Fund under the No 09.3.3--LMT--K--712 "Development
of Competences of Scientists, other Researchers and Students through Practical Research Activities"
measure (Project No. 09.3.3--LMT--K--712--23--0235).


\end{document}